\documentclass[multphys,vecphys]{svmult}

\usepackage{makeidx}         
\usepackage{graphicx}        
\usepackage{multicol}        
\usepackage[bottom]{footmisc}

\makeindex             

\newcommand{\microns}{\ensuremath{\mu\mbox{m}}}

\begin{document}

\title*{VITRUV - Imaging close environments of stars and galaxies with
  the VLTI at milli-arcsec resolution}
\titlerunning{The VITRUV instrument} 

\author{
     Fabien Malbet \inst{1}
\and Jean-Philippe Berger \inst{1}
\and Paulo Garcia \inst{2} 
\and Pierre Kern \inst{1}
\and Karine Perraut \inst{1} 
\and Myriam Benisty \inst{1}
\and Laurent Jocou \inst{1}
\and Emilie Herwats \inst{1,3}
\and Jean-Baptiste Lebouquin \inst{1}
\and Pierre Labeye \inst{4}
\and Etienne Le Coarer \inst{1}
\and Olivier Preis \inst{1}
\and Eric Tatulli \inst{1}
\and Eric Thi\'ebaut \inst{5}
}
\authorrunning{F. Malbet et al.}

\institute{
  Laboratoire d'Astrophysique de Grenoble, BP 53, F-38041 Grenoble
  cedex 9, France \texttt{Fabien.Malbet@obs.ujf-grenoble.fr} 
  \and %
  Centro de Astrof\'{\i}sica da Universidade do Porto, Rua das Estrelas,
  4150-762 Porto, Portugal 
  \and %
  Universit\'e de Li\`ege, Li\`ege, Belgique
  \and %
  CEA-LETI, Grenoble, France
  \and %
  Centre de Recherche en Astrophysique de Lyon, Lyon,
  France
}
%
%
\maketitle

\begin{abstract}
  The VITRUV project has the objective to deliver milli-arcsecond
  spectro-images of the environment of compact sources like young
  stars, active galaxies and evolved stars to the community. This
  instrument of the VLTI second generation based on the integrated
  optics technology is able to combine from 4 to 8 beams from the VLT
  telescopes. Working primarily in the near infrared, it will provide
  intermediate to high spectral resolutions and eventually
  polarization analysis. This paper summarizes the result from the
  concept study led within the Joint Research Activity \emph{advanced
    instruments} of the OPTICON program.
\end{abstract}

\section{Introduction}

The VLT interferometric facility is unique in the world, since it
offers giant 8m telescopes, 2m auxiliary telescopes and the necessary
infrastructure to combine them. With four 8m unit telescopes (UTs)
equipped with adaptive optics systems, four 1.8m auxiliary relocable
telescopes (ATs) equipped with tip-tilt correction, a maximum
separation of 130m for UTs and 200m for ATs, 6 available delay lines,
slots foreseen for 2 more ones, a dual feed capability (PRIMA) and a
complete system control, the VLTI is the best site to propose the
first optical interferometer to deliver routinely aperture synthesis
images like the millimeter wave interferometers are already doing for
more than 10 years. The quality of the images will be as good as the
ones delivered by the IRAM Plateau de Bure Interferometer with six 15m
antennas and a maximum baseline of 500m at 1-3mm. The VLTI will be for
a long time the only facility with 10m class telescopes able to
provide images with 1mas angular resolution in optical wavelength.

We propose a second generation instrument for the VLTI, called VITRUV,
aimed at taking the best profit of the imaging capability of the
array, especially within the PRIMA framework. The science objectives
of VITRUV are focused on the kinematics and morphology of compact
astrophysical objects at optical wavelengths like the environment of
AGN, star forming regions, stellar surfaces and circumstellar
environments. The instrument will deliver aperture synthesis images
with spectral resolution as the final data product to the astronomer.

The specifications can be summarized as:
\begin{itemize}
\item beam combiners for 4T and 8T operation,
\item a temporal resolution of the order of 1 day,
\item 2 or 3 spectral resolutions from 100 to 30000,
\item image dynamics from 100 to 1000,
\item a field of view up to1 arcsec
\item initial wavelength coverage from 1 to 2.5 microns that could be extended from 0.5 to 5 microns. 
\end{itemize}
The technology that is contemplated at this stage is integrated optics
because it offers simplicity, stability especially for phases,
operational liability, and high performances. This technology has been
already successfully validated on the 3 telescope IOTA interferometer
where the system routinely delivers visibilities and closure phases
for 3 baselines \cite{Monnier2004,Kraus2005} and on the VLTI to replace
the fiber coupler of VINCI \cite{LeBouquin2004a}.

\section{Science objectives}

The science cases definition methodology was to concentrate in a few
fields where VITRUV can make a substantial contribution, without being
fully exhaustive.  We list here the four science cases for the VITRUV
concept where significant advance can be achieved:
\begin{enumerate}
\item Studying the formation of stars and planets by direct
  spectro-imaging of their inner disk regions, from the orbits of
  Mercury to Neptune. 
\item Imaging the magnetic and convective hallmarks of stellar
  surfaces.
\item Connecting the geometries of the close environments of evolved
  stars with their progenitors.
\item Probing the close environment of active gaglactic nuclei and
  supermassive black holes.  
\end{enumerate}
Further details are given in the \emph{VITRUV Science Cases} in this
volume \cite{Garcia2005}. The top level technical requirements has been
deduced from these science requirements:
\begin{itemize}
\item One night imaging capability thanks to the largest simultaneous
  (u,v) plane coverage while combining simultaneously between 4 and 8
  beams (can be downgraded to 6)
\item Spectral resolution from 100 up to 30,000
\item Sensitivity and spectral resolution requires phase stabilization
  (fringe tracker) and eventually dual beam referencing (PRIMA). 
\end{itemize}

VITRUV has been
designed to be the instrument for phase reference imaging (PRIMA) with
the telescopes available on the site. In addition it can perform phase
closure imaging in the case where there is no adequate reference star
in the field. Therefore it is important to consider VITRUV in the
PRIMA development scheme although this service is not mandatory for
relatively bright targets (up to $K=11-13$ with the UTs).

\section{VITRUV concept}

\begin{figure}[t]
  \centering
  \includegraphics[width=0.6\hsize]{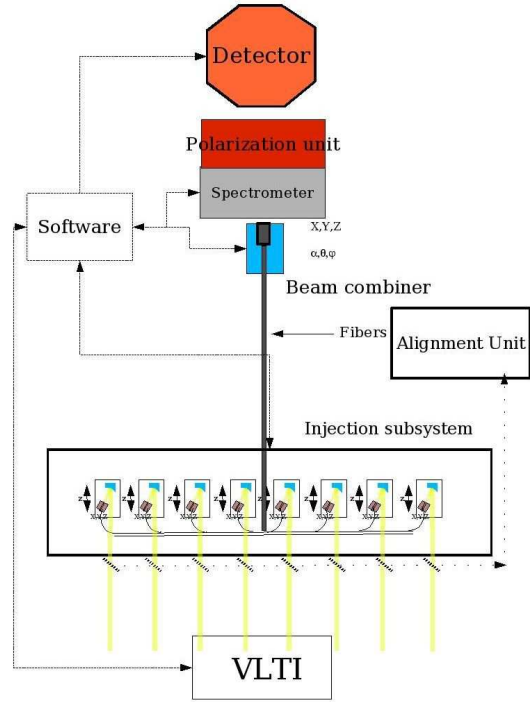}
  \caption{Vitruv concept}
  \label{fig:concept}
\end{figure}

VITRUV is a non-direct imaging instrument. It measures electromagnetic
field complex coherence at different spectral resolutions, different
polarization states for a maximum of 28 baselines (8
telescopes). Visibilities, phases and closure phases are the raw
observables.  It is a single-mode instrument with field of view
capability of a few Airy disks. The data product is the reconstructed
spectral image cube at all wavelengths. 

VITRUV is designed to simplify operation as much as possible. A
self-aligned integrated optics (IO) beam combiner is at the heart of
the instrument concept. At this stage of the study we plan to have
four beam combiners to cover the J, H, K bands: 
\begin{itemize}
\item optimized 4-way beam combiner for J/H band
\item 5 to 8 way beam combiner for J/H band
\item optimized 4-way beam combiner for K band
\item 5 to 8 way beam combiner for K band,
\end{itemize}
and two additional ones to extend to R/I and L bands. Each 8-way beam
combiner has a maximum of 8 inputs and can be used with whatever
combination of telescopes is available (i.e. from 4 to 8). In order to
limit the size of the instrument, specific injection modules under
development at LAOG will allow selecting which combiner is in use at a
given time. VITRUV concept has a spectral resolution
capability with an optional polarization state capability which will
be traded-off between astrophysical goals and instrument complexity.

\begin{figure}[t]
  \centering
  \includegraphics[width=0.9\hsize]{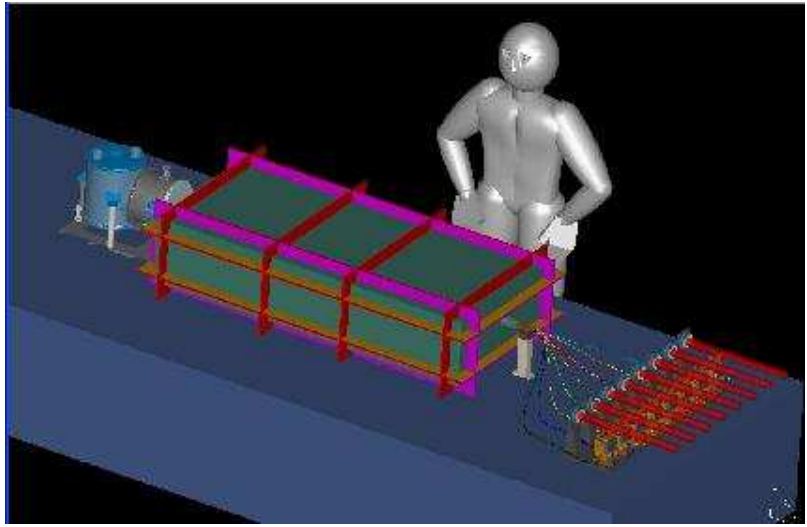}
  \caption{General view of the VITRUV instrument}
  \label{fig:3d}
\end{figure}
The VITRUV instrument (see Fig.~\ref{fig:3d}) can be described by 6
subsystems as follows.

\subsection{Injection subsystem}

This system has two functions: (1) to inject VLTI beams into VITRUV
beam combiner fibers, and (2) to select the chosen combiner thanks to
a motorized translation. Focus will be made thanks to off-axis
parabolae in order to discard chromaticity effects. At the focus of
each parabola a miniaturized fiber positioner
\cite{Preis2004} holds N fibers, N being the number of beam
combiners (4 in the JH/K only version). Fiber numerical apertures will
be chosen in order to optimize an average coupling covering the VITRUV
band pass.  A 1D translation allows us to select between the N fibers.
A miniaturized fiber positioner allows on-sky flux optimization. Each
injection module is located on a translation stage designed to
equalize optical paths for internal fringe acquisition purposes.

\subsection{Integrated optics beam combiners}  

The optical beam combining function will be achieved thanks to
integrated optics (IO) technologies that allow embedding single mode
optical circuits in glass or silica chips. As of today we plan to use
two different types of beam combiners, one for 4-way beam combination
and one for 5 to 8-way beam combination. The beam combination
principle (coaxial or multiaxial) is different between the two in
order to maximize the signal-to-noise ratio \cite{LeBouquin2004b}.


For the JH/K only version of VITRUV the 4 beam combiners are stacked
together (or integrated in the same chip, not defined as of today) on
a remotely controlled motorized 6-axes positioner which allows us to
position the beam combiner output in front of the detector. Only one
beam combiner is operational at the same time, selected by the
injection module (i.e. on wavelength band) and, if required, by the
combiner positioner. 

The total number of degrees of freedom for VITRUV is $8+56 = 64\,\mbox{dof}$. 

\subsection{Spectrometer}  

Preliminary design of the spectrometer will be mainly constrained by
science case requirements and technical issues. It will include
motorized axes in order to select spectral resolution mode. IO
combiners add a supplement interesting property in the sense that the
chip can act as an entrance slit for a spectrometer and does not
require optical anamorphous transform.

\subsection{Detector}  

The detector choice is still at its premises. The JHK detector will be
a nitrogen cooled array with low readout noise but fast enough to be
able to read quadrants and sample spectrally dispersed fringes within
the atmosphere perturbations eventually compensated by the fringe
tracker.

Performant detectors can also now be found for the R/I part of the
spectrum and L band.  The proposed visible extension requires an
additional development for a low noise fast read out detector or
photon-counting detector, with performances close to the performances
of an AO visible wavefront sensor (see OPTICON development in JRA2
based on an EEV detector 288x288). 

\subsection{Software}  

Software development is required to ensure control of all VITRUV
motorized elements, camera readout and data acquisition, interface
with VLTI software, data reduction, image reconstruction. We do not
expect that the VITRUV instrument control software to be very
different from the AMBER software except for the number of input
beams. We plan to use the maximum of the heritage from the AMBER
software.

Like for AMBER the data reduction software (DRS) will be part of the
package, but in addition we are working in collaboration with the
Jean-Marie Mariotti Center (JMMC, see \texttt{http://mariotti.fr}) to provide
image reconstruction software to provide reconstructed images to the
users.

\subsection{Polarization control} 

As of today two VITRUV instrumental modes that will deal with
polarization issues are contemplated: \emph{polarization split at the
  output} recording two linear polarizations and relax constrains due
to the use of birefringent waveguides. With proper calibration this
mode should allow to provide information on the degree of linear
polarization of the source; \emph{a full polarization analyzing
  module} (should the science case demonstrate its importance) where
linear and circular differential polarization states will be measured.

\section{VITRUV within the VLTI infrastructure}

VITRUV is an instrument that requires a full and operational VLTI
infrastructure. 

UT operation requires adaptive optics capability which is already
available with the MACAO systems. The availability of AO systems on AT
would significantly improve the capabilities of the instrument, mainly
to reach shorter wavelengths ideally down to the R band. 

The considered concept allows long exposure acquisition requested for
the high spectral resolution mode. Therefore the sensitivity and the
spectral coverage of the instrument will be highly improved with
fringe tracking capabilities\footnote{This study can be developed in
  parallel if this fringe tracking facility exists, but if not then it
  should be added to the project.}. In addition with the PRIMA
facility if adapted to all telescope subsystems, the imaging
capability can be pushed towards fainter sources with a bright
reference nearby.

\section{Expected performances}

The performances of the VITRUV instrument depend on the sensitivity of
the fringe tracker. We assume here that this fringe tracker can work
up to K=11. Since the fringe tracker is a low pass filter, we assume
that the piston correction is almost perfect over a few seconds.




\begin{figure}[t]
  \centering
  \includegraphics[width=0.3\hsize]{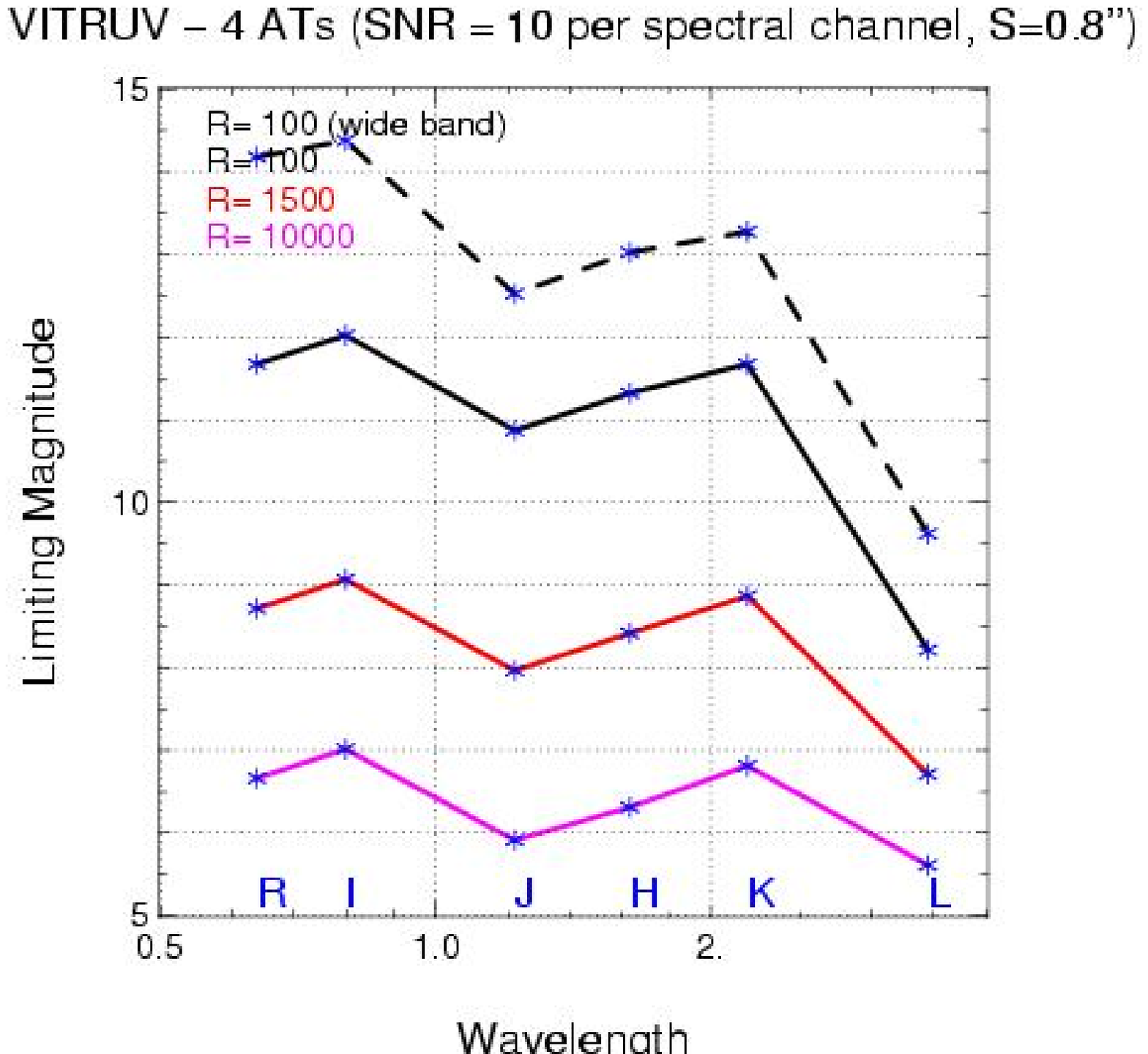}
  \includegraphics[width=0.3\hsize]{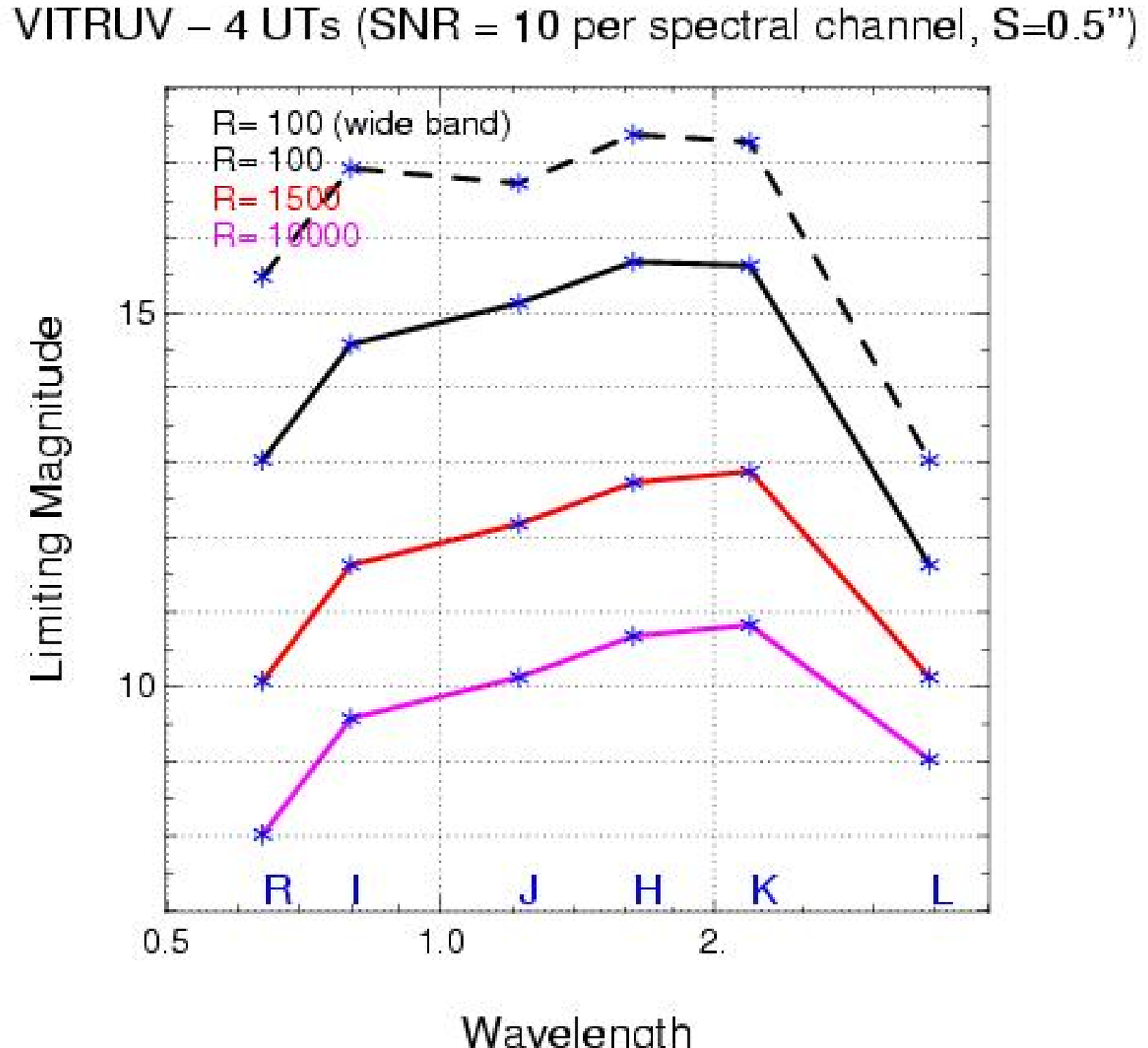}
  \includegraphics[width=0.3\hsize]{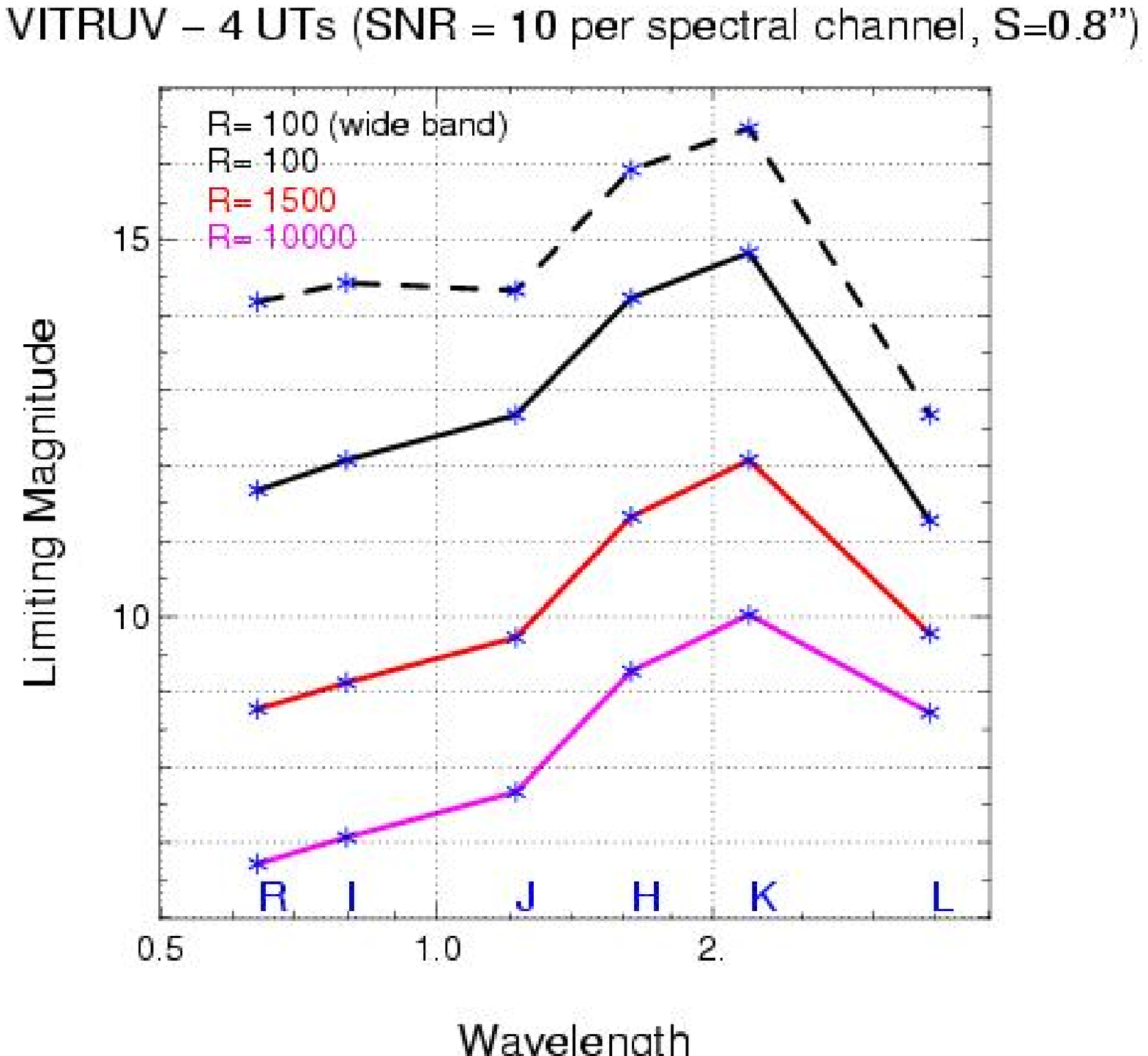}
  \caption{Limiting magnitude through the VITRUV wavelength coverage.
    Left panel with 4 ATs, middle panel with 4 UTs and an average
    seeing of $0.8"$, and right panel with 4 UTs and an exceptional
    seeing of $0.5"$.}
  \label{fig:snr}
\end{figure}

We have computed SNR curves with a perfect external PRIMA fringe
tracker. The limiting magnitude of this PRIMA fringe tracker should be
the one computed by AMBER in the low resolution mode, i.e. $K=11-13$
depending on average seeing conditions. We assumed in our calculation
that the AO guide star is bright and is $V=5$. The conditions remain
about the same up to $V=13$. Since the fringes are stabilized, we used
a 100s elementary exposure time. We have not yet fully investigated
the dual feed option in simulations (losses due anisoplanetism).

For imaging the requested visibility accuracy does not need to be very
stringent. In mm radio interferometry, maps are produced with 10\%
visibility errors. More important is the (u, v) coverage. A typical
visibility accuracy below 1\% and phase accuracy of the order of 1
degree is sufficient and already achieved
\cite{Berger2001,LeBouquin2004a,Monnier2004,Kraus2005}.

The field of view of VITRUV is fundamentally limited by the FOV
accessible by the injection fibers, i.e. in $K$ 250 mas with the ATs and
60 mas with the UTs. Most of the science which is contemplated focuses
on compact objects within this limit. However like for radio
interferometry, we know that we can extend the FOV by performing
mosaicing (Tatulli, 2004).

We focus the project on objects with a bright central reference, but
no so bright that we need to cancel it. Basically a dynamic range
between the faintest and the brightest features in the image between
100 and 1000 are achievable. We have started to build a software-based
end-to-end simulator of VITRUV which allows us to investigate this
issue with more accuracy.






\section{Project management}

At this stage, it is difficult to predict an accurate evaluation of
the resources required for the project.

\textbf{Cost}.  A first rough estimation for an instrument operable in
the 1-2.5\microns\ range leads to a cost of 1.2\,MEuros. The
additional cost to extend the wavelength range (0.6 - 4\,\microns) is
1.1\,MEuros. This extension includes an additional camera for the
visible and some technological development for the integrated optics
components.

\textbf{Manpower}. The 1-2.5\,\microns\ instrument requires a manpower
support of 56 FTE for its design and construction. The required
additional manpower is 22 FTE for a 0.6-4\,\microns\ extension.

\textbf{Schedule}.  The required time to achieve the full
manufacturing of the 1-2.5\,\microns\ instrument is 3.5 years.  Six
additional months will be required to achieve the extended version,
taking into account that main additional developments for the proposed
extension will be carried out in parallel during the whole project.

\textbf{Collaborations}. For the moment, a formal consortium has not
yet been established. We are waiting for the end of the selection
process to start working on this issue.  We are very open to various
forms of collaborations.

\bibliographystyle{springer}
\bibliography{malbet-vitruv}

\end{document}